\documentclass[final,5p,times,twocolumn]{elsarticle}
\usepackage[utf8]{inputenc}
\usepackage[T1]{fontenc}
\usepackage{graphicx}
\usepackage{longtable}
\usepackage{wrapfig}
\usepackage{rotating}
\usepackage[normalem]{ulem}
\usepackage{amsmath}
\usepackage{amssymb}
\usepackage{capt-of}
\usepackage{hyperref}
\usepackage{siunitx}
\date{}
\begin{document}

\begin{frontmatter}
\title{The Mu2e Experiment --- Searching for Charged Lepton Flavor Violation}
\author[addP]{M. T. Hedges\corref{cor1}}
\author{on behalf of the Mu2e Collaboration}
\address[addP]{Department of Physics and Astronomy, Purdue University, 525 Northwestern Avenue, West Lafayette, IN 47907, USA}
\cortext[cor1]{Corresponding author: hedges7@purdue.edu}
\begin{abstract}
The Mu2e experiment will search for a Standard Model violating rate of neutrinoless conversion of a muon into an electron in the presence of an aluminum nucleus. Observation of this charged lepton flavor violating process would be an unambiguous sign of new physics. Mu2e will improve upon previous searches for this process by four orders of magnitude. This requires the world's highest-intensity muon beam, a detector system capable of efficiently reconstructing the 105 MeV/c conversion electron signal, and minimizing sensitivity to background events. A pulsed 8 GeV proton beam strikes a target, producing pions that decay into muons. Beam outside the pulse must be suppressed to $< 10^{-10}$ to reduce beam-related backgrounds. The muon beam is guided from the production target along the transport system and onto the aluminum stopping target. Conversion electrons leave the stopping target and propagate inside a solenoidal magnetic field to the tracker and electromagnetic calorimeter. The tracker is a system of straw tube panels filled with Ar/CO$_2$ at 1 atm that tracks particles inside of a solenoidal B-field and measures their momenta with $\sim100$ keV/$c$ resolution to resolve signal events from decay-in-orbit backgrounds. The CsI calorimeter provides $E/p$ and is used to seed the track reconstruction algorithm with $\sigma_E /E \sim{}10\%$ and $\sigma_t<500~\rm{ps}$. Additionally, a novel cosmic ray veto with greater than 99.99\% efficiency brings the expected number of background events to fewer than one over three years of running. To normalize the experiment, the stopping target monitor measures the rate of capture photons from muons incident on the stopping target by using a system of high-purity germanium and lanthanum bromide scintillators.
\end{abstract}
\end{frontmatter}

\section{Introduction}
The Standard Model of particle physics classifies fundamental fermions into six quark flavors and six lepton flavors. Of these fermions, only the charged leptons have never experimentally shown evidence of flavor violation. However, the discovery of massive neutrinos provides a mechanism for charged lepton flavor violation (CLFV) at loop level, but is highly suppressed to unobservably small levels. Thus, any observation of CLFV would unambiguously indicate physics beyond the Standard Model.

One particularly promising experimental search for CLFV focuses on detection of direct conversion of a muon to an electron in the Coulomb field of a nucleus, or $\mu^-N - e^-N$ conversion. This process produces a monoenergetic \emph{conversion-electron} (CE) with energy given by: $$ E_{CE} = m_\mu{}c^2 - E_b - E_\text{recoil} $$
where $E_b$ is the binding energy of the muon in the $1S$ orbit and $E_\text{recoil}$ is the energy of the recoiling nucleus.

\section{The Mu2e experiment}
The Mu2e experiment \cite{mu2e-tdr} will search for $\mu^-N - e^-N$ by measuring \(R_{\mu e}\), defined as the ratio of the rate of $\mu^-N - e^-N$ conversion to the rate of muon capture using an aluminum target nucleus. Specifically, Mu2e will measure \(R_{\mu e}\) on an Al target given by: $$R_{\mu e} = \frac{\mu^- + A(Z,N) \to e^- + A(Z,N)}{\mu^- + A(Z,N) \to \nu_\mu + A(Z-1,N)}$$ with a 5\(\sigma\) discovery potential of \(R_{\mu e} > 2 \times 10^{-16}\) or a corresponding upper limit of \(R_{\mu e} < 8 \times 10^{-17}\) (90\% CL). This sensitivity is four orders of magnitude beyond the current bounds set by SINDRUM II which measured \(R_{\mu e} < 7 \times 10^{-13}\) on Au \cite{sindrumII}.

\begin{figure*}
\begin{centering}
\includegraphics[width=\textwidth]{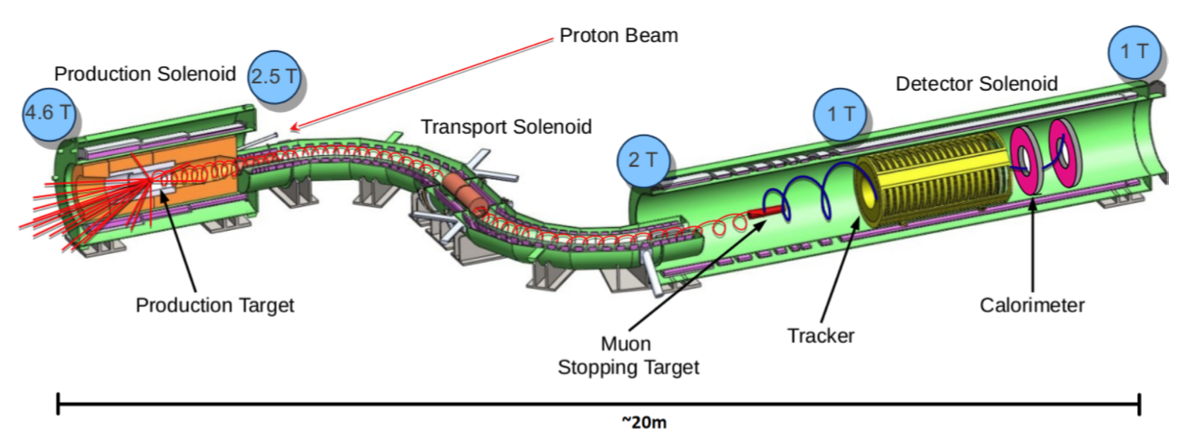}
\caption{\label{fig:mu2e-full}The Mu2e experiment.}
\end{centering}
\end{figure*}

For $\mu^-N - e^-N$ in aluminum, the CE signal is a monoenergetic electron of $E_{CE} = 104.9~\rm{MeV}/c$. Mu2e must detect this signal in the presence of both \emph{intrinsic} backgrounds such as muon Decay-In-Orbit (DIO) and cosmic-ray events, and \emph{beam-induced} backgrounds such as antiproton annihilation and radiative pion capture (RPC) in the muon stopping target. These two types of backgrounds drive the design of the Mu2e experiment, which is shown in Fig. \ref{fig:mu2e-full}.


\begin{figure}
    \centering
    \includegraphics[width=\columnwidth]{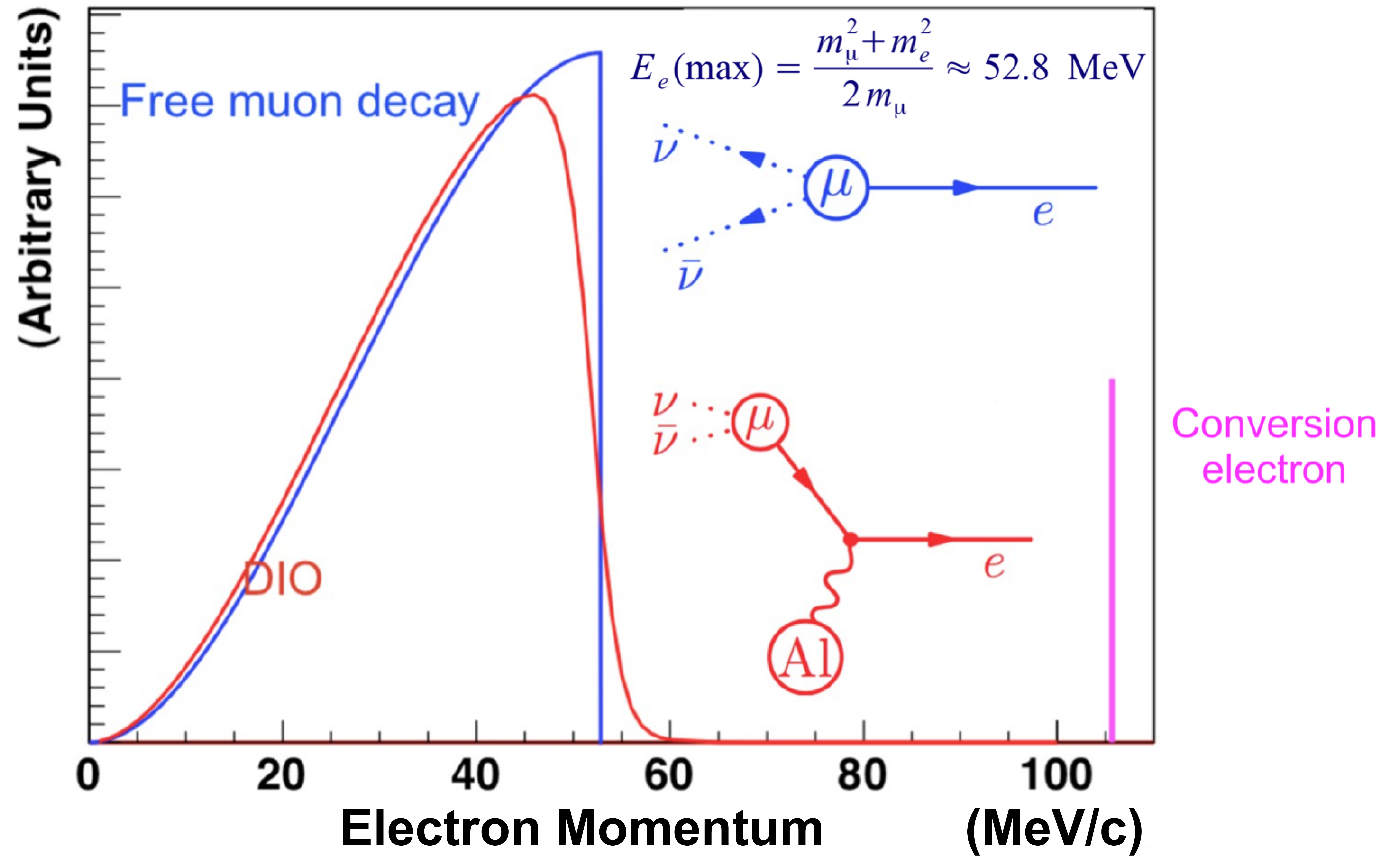}
    \caption{Qualitative representation of electron momentum spectrum due to free muon decay (blue), muon decay-in-orbit (red), and conversion electrons (magenta).}
    \label{fig:mu2e-dio-ce}
\end{figure}

DIO events are the primary intrinsic background of concern. Muons incident on the aluminum stopping target can decay while in atomic orbit to an electron and two neutrinos, just as a free muon. While the DIO electron momentum spectrum is similar to that of free muon decay, the presence of the aluminum nucleus can cause a recoil and results in the kinematic endpoint of DIO electrons to be equivalent to the momentum of the CE signal. This is demonstrated graphically in Fig. \ref{fig:mu2e-dio-ce}. A full calculation of the DIO spectrum in aluminum can be found in Ref. \cite{2016_Szafron_DIO_LL_PhysRev}.

\begin{figure}
\centering
\includegraphics[width=0.75\columnwidth]{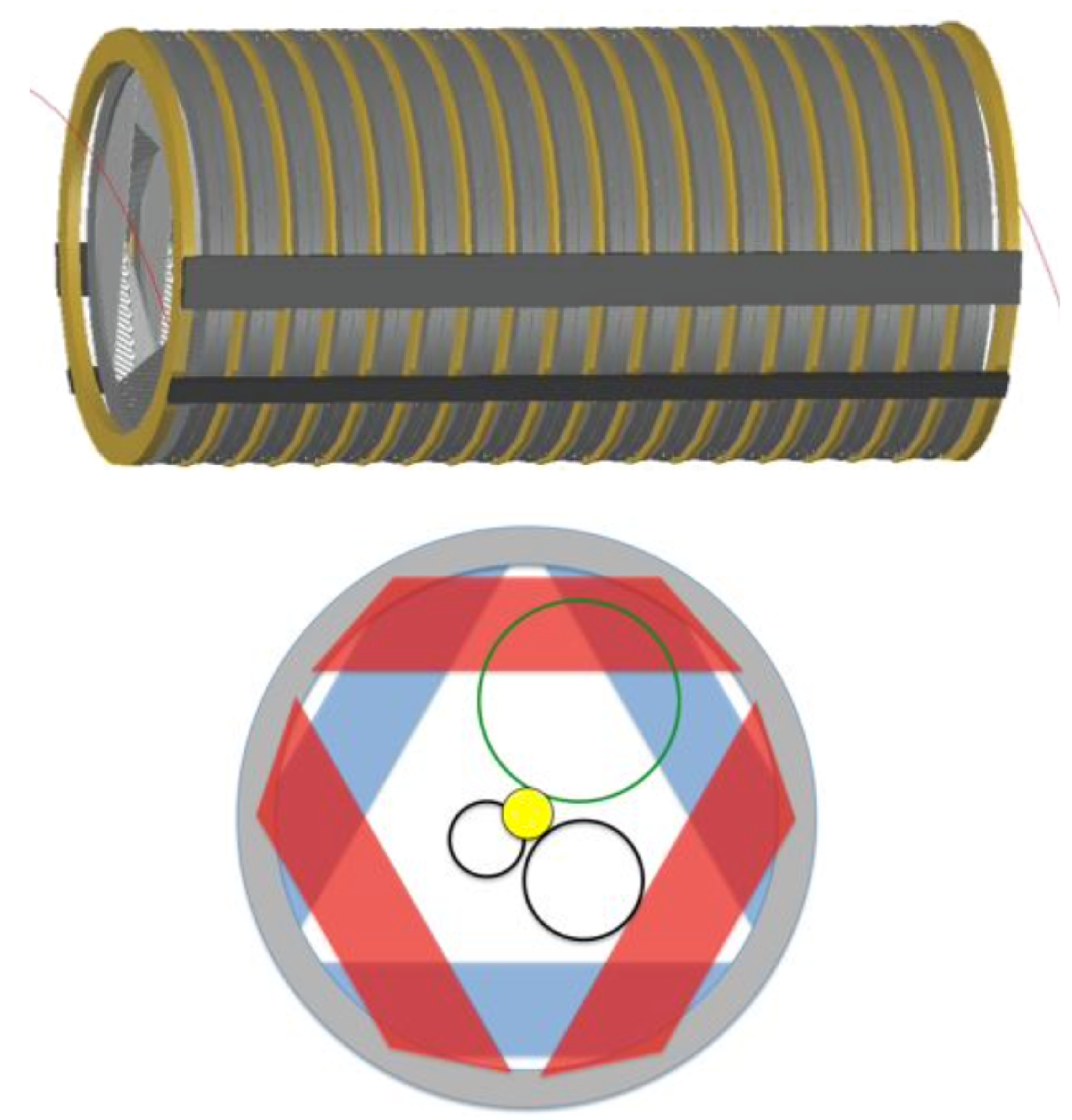}
\caption{\label{fig:mu2e-tracker}(top) The Mu2e straw tube tracker and (bottom) the acceptance of the annular design to the radii of 105 MeV/$c$ conversion electrons (green) and lower momentum backgrounds (black) emitted from the stopping target.}
\end{figure}

Electrons emitted from the stopping target traverse a solenoidal magnetic field into the Mu2e tracker. Resolving the CE signal from the DIO background requires the tracker to have excellent momentum resolution and track reconstruction efficiency. This is obtained by utilizing a low-mass straw-tube tracking system consisting of 21,600 mylar straws of 15 \si{\micro\meter} thickness arranged into 36 panels. The straws are filled with 1 atm of 80:20 Ar:CO$_2$ with an applied voltage of 1450 V. The tracker is expected to provide a momentum resolution of $\sim{}100~\rm{keV}/c$ at the CE momentum of $\sim{}105~\rm{MeV}/c$. To maximize track reconstruction efficiency and to minimize effects from the large currents produced from low-momentum DIO events, the tracker utilizes an annular design wherein an inner radius of 38 cm remains uninstrumented. This design, shown in Fig. \ref{fig:mu2e-tracker}, drastically reduces acceptance to lower momentum electrons while maintaining acceptance to conversion electrons.

\begin{figure}[htbp]
\centering
\includegraphics[width=0.99\columnwidth]{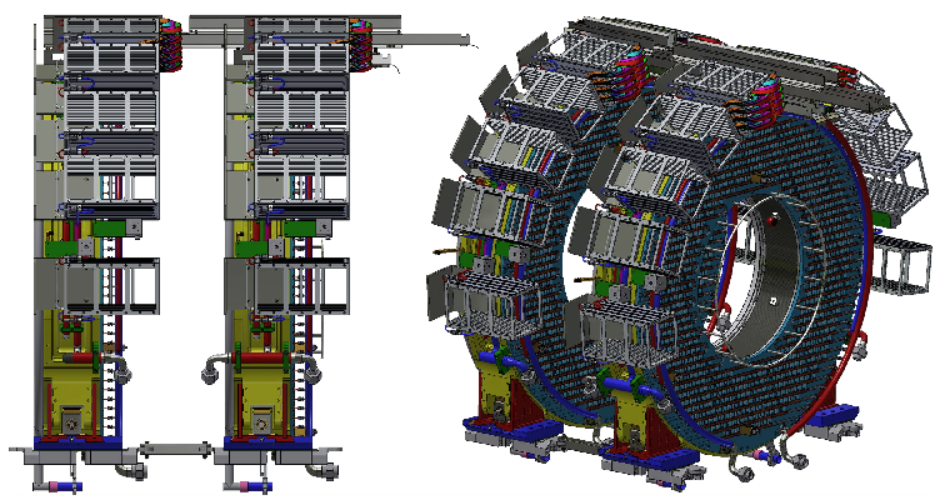}
\caption{\label{fig:mu2e-calorimeter}The Mu2e CsI calorimeter.}
\end{figure}

The calorimeter, shown in Fig. \ref{fig:mu2e-calorimeter}, consists of two disks of 674 pure CsI crystals arranged in the same annular design as the tracker. Each CsI crystal measures 34\texttimes{}34\texttimes{}200 mm\textsuperscript{3} and is read out by 2 SiPMs. The calorimeter provides \(E/p\) for particle identification with expected  \(\sigma_E/E\) of order 10\% for conversion electrons, as well as timing resolution of \(\sigma\)\textsubscript{t} < 500 ps, which seeds the Mu2e track reconstruction algorithm. Prototype modules consisting of an array of $3 \times 3$ crystals have been tested utilizing a variable energy electron beam (22 -- 120 MeV) have demonstrated an energy resolution of $\sim$7\% and a time resolution of better than $\sim$230 ps \cite{Atanov_2018}.

\begin{figure}
\centering
\includegraphics[width=0.99\columnwidth]{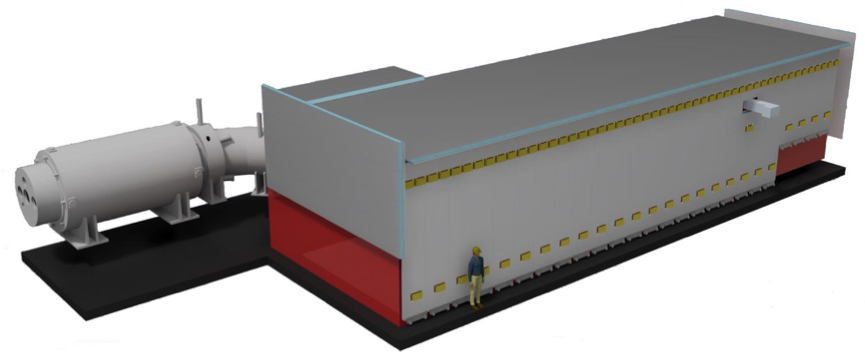}
\caption{\label{fig:mu2e-crv}The Mu2e Cosmic Ray Veto.}
\end{figure}

The remaining significant intrinsic background arises from cosmic rays, which can enter the experimental apparatus and produce CE-like events. Simulations have indicated a base rate of order 1 CE-like event per day of running. To reject these events, Mu2e uses a Cosmic Ray Veto (CRV) which surrounds the Detector Solenoid that houses the muon stopping target, the tracker, and the calorimeter. The CRV consists of over 5000 extruded polystyrene scintillators arranged in four overlapping layers. Scintillation light is collected by wavelength shifting fibers and detected by SiPMs. This design, shown in Fig. \ref{fig:mu2e-crv}, is expected to reject cosmic ray events with over 99.99\% efficiency using triple coincidence of the scintillator layers.

\begin{figure*}
\begin{centering}
\includegraphics[width=\textwidth]{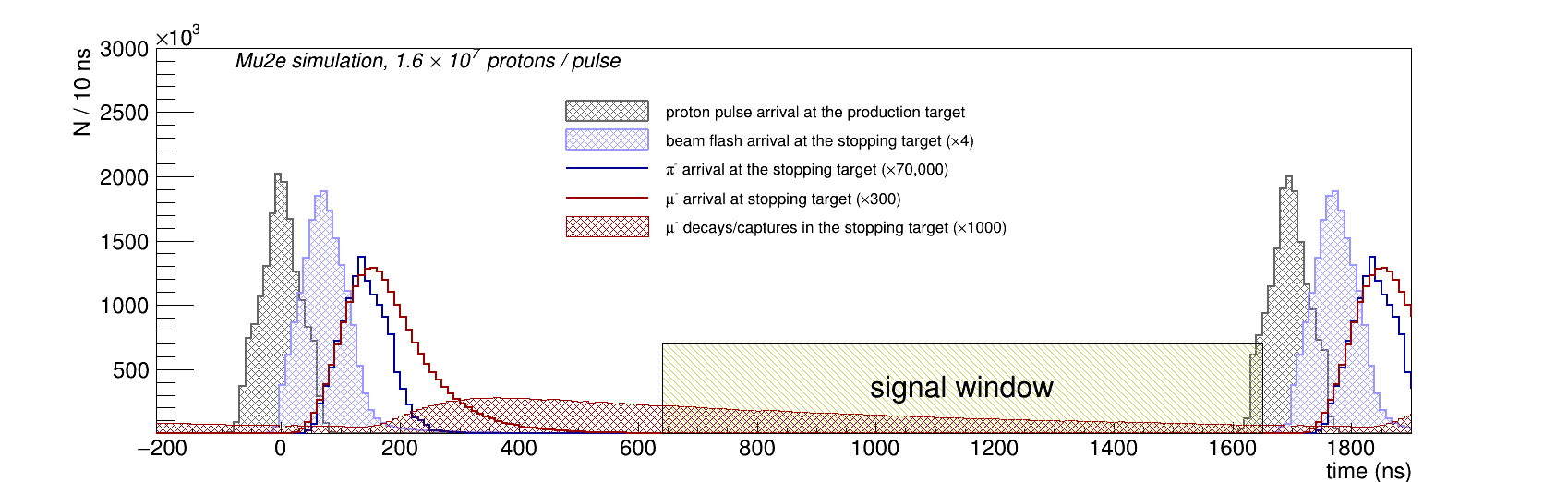}
\caption{\label{fig:mu2e-livegate}Timing distributions for the CE signal and various backgrounds, with the signal window shown at times after the reduction of prompt backgrounds and before the arrival of backgrounds from the next pulse.}
\end{centering}
\end{figure*}

Beam-induced backgrounds can also limit sensitivity to CE signal. The muon beam for Mu2e originates from a pulsed proton beam incident on the tungsten pion Production Target. A single pulse consists of \(\sim\)4 \texttimes{} 10\textsuperscript{7} protons at 8 GeV, with a pulse duration of \(\sim\)250 ns and period of 1695 ns---longer than the muonic-Al lifetime of 864 ns. Pions emerge from the production target and decay to muons, which are then transferred down the Transport Solenoid and onto the Al muon Stopping Target inside of the Detector Solenoid. The pulsed beam allows the stopped muons inside the stopping target to decay during the time between sequential pulses. This allows blinding the experiment until after prompt beam backgrounds decay to appropriate levels. This is shown in Fig. \ref{fig:mu2e-livegate}, which shows the intensity versus time for various sources of detectable electrons. As can be seen, using a delayed signal window drastically reduces sensitivity to beam-induced backgrounds.

\begin{figure}
\centering
\includegraphics[width=0.99\columnwidth]{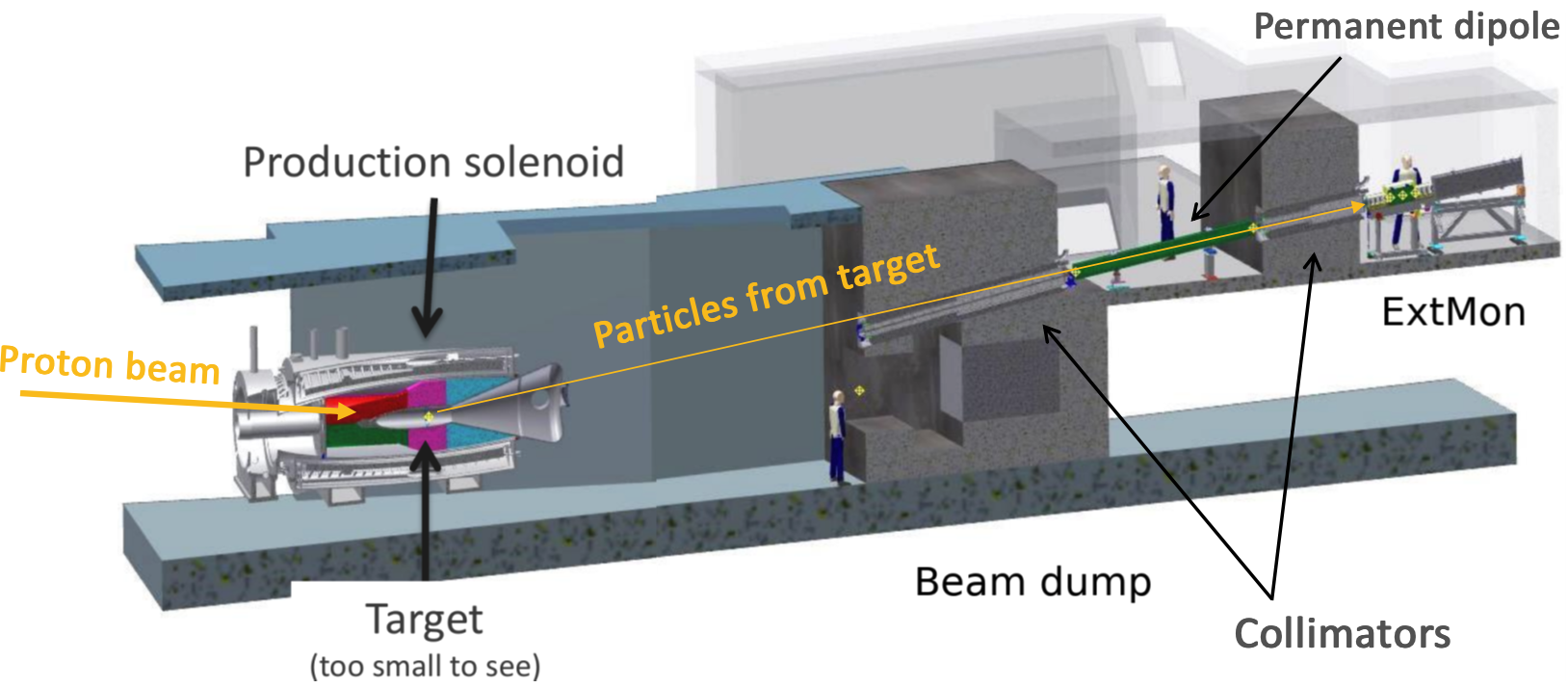}
\caption{\label{fig:mu2e-extmon}The Mu2e Extinction Monitor, shown relative to the production target.}
\end{figure}

Achieving this level of reduction in beam-induced backgrounds requires a minimal level of beam particles to arrive at the muon stopping target within the signal window. We require this inter-pulse \emph{extinction}---defined as the ratio of out-of-time beam intensity to in-time beam intensity---to be less than 10\textsuperscript{-10}. The Extinction Monitor pixel telescope, shown in Fig. \ref{fig:mu2e-extmon}, consisting of eight sequential planes with 20 FE-I4B pixel chips, measures the extinction by tracking protons scattered downstream off of the Production Target both in- and out-of-time with the proton pulses.

\begin{figure}
\centering
\includegraphics[width=0.99\columnwidth]{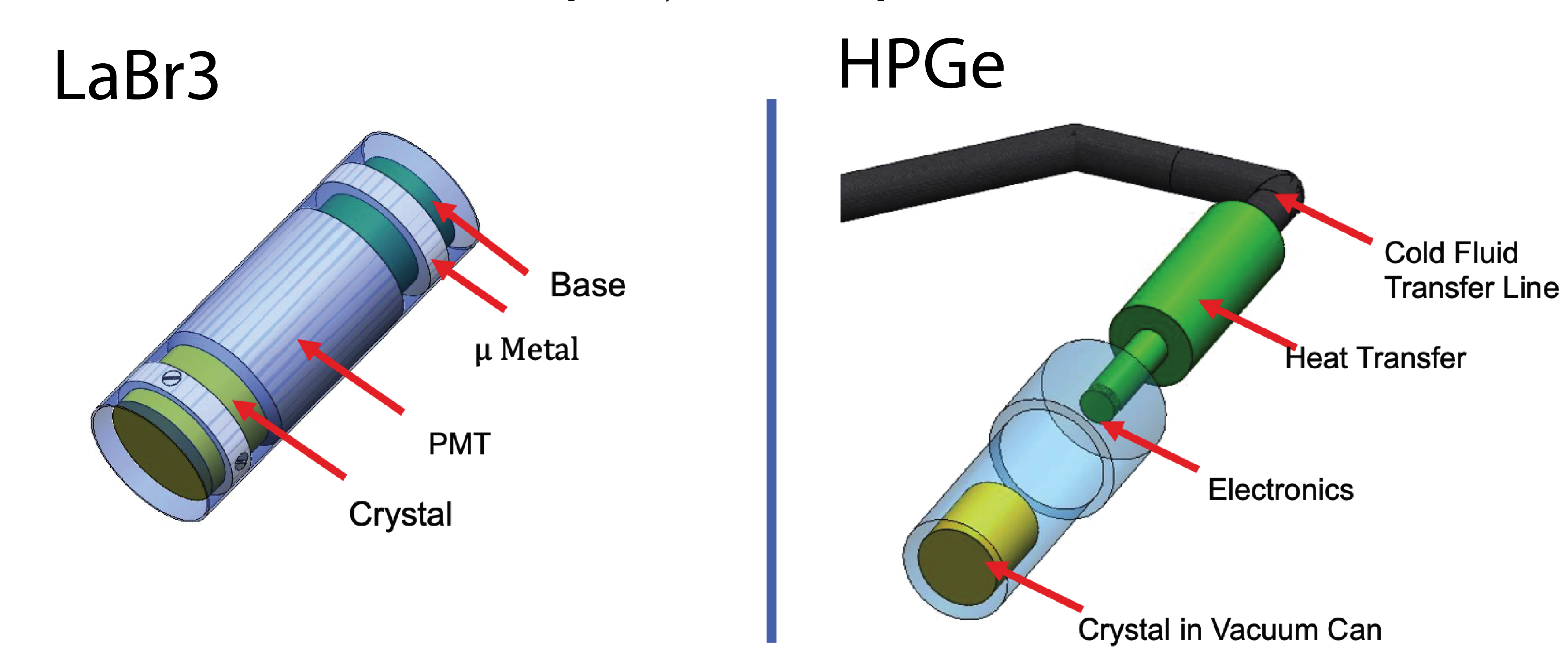}
\caption{\label{fig:mu2e-stm}The LaBr$_3$ (left) and HPGe (right) detectors used in the Mu2e Stopping Target Monitor.}
\end{figure}

Finally, the Stopping Target Monitor provides the absolute normalization by measuring the gamma spectrum from atomic capture of muons on the Al stopping target using both LaBr\textsubscript{3} and high-purity Ge detectors. A mockup of the LaBr$_3$ and HPGe sensors is shown in Fig. \ref{fig:mu2e-stm}.

\section{Mu2e Run 1}
\begin{figure}
\centering
\includegraphics[width=0.99\columnwidth]{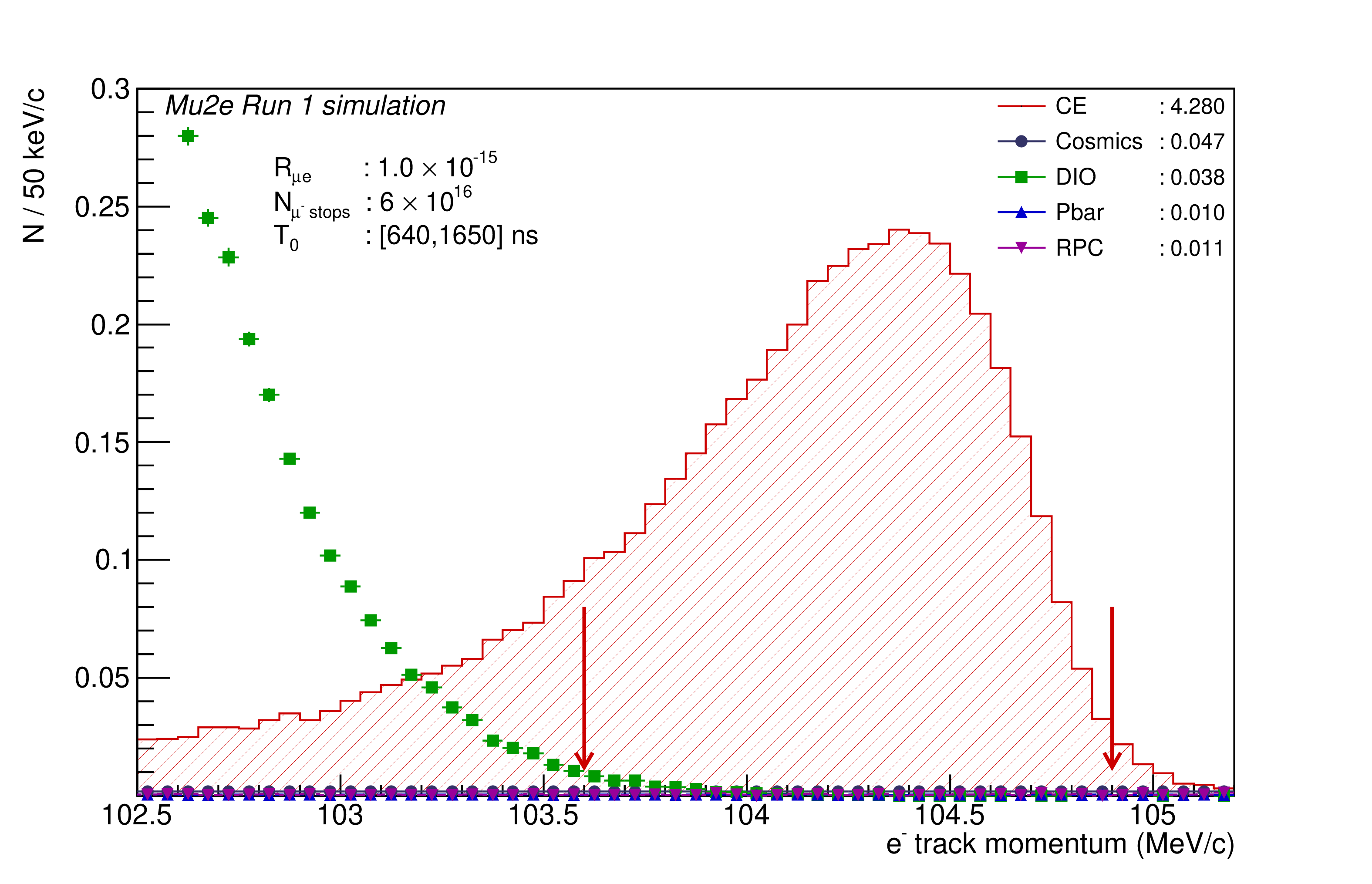}
\caption{\label{fig:mu2e-su2020}Expected momentum spectra from conversion electrons in the presence of backgrounds during Mu2e Run I.}
\end{figure}

The Mu2e experiment is poised to begin Run 1 of physics data with muon beam during the middle of this decade. Through a dedicated simulation campaign utilizing expected detector and accelerator performance during the expected duration of Run 1, we expect a discovery potential of $R_{\mu e} > 1 \times 10^{-15}$, with a standalone, dedicated publication anticipated this year. The simulated momentum spectrum for signal and background components is shown in Fig. \ref{fig:mu2e-su2020}, and the yields for expected background contributions are shown in Table \ref{tbl:mu2e-run1}. In the absence of signal in Run 1, the corresponding limit will be improved to  $R_{\mu e} < 6 \times 10^{-16}$ (90\% CL), corresponding to 1000 times better than the current world limit. Run 2, with an anticipated start by the end of the decade, will provide an additional factor of 10 in improvement beyond Run 1, resulting in an overall $10^4$ improvement in sensitivity.

\begin{table}[htbp]
\caption{\label{tbl:mu2e-run1}Expected yields of backgrounds in Mu2e Run 1 with statistical and systematic uncertainties.}
\centering
\begin{tabular}{cc}
 & Expected yield \\
\hline
Cosmics & $0.048 \pm 0.01 \pm 0.010$\\
DIO & $0.038 \pm 0.002~^{+0.026}_{-0.016}$\\
RPC & $0.011 \pm 0.002~^{+0.001}_{-0.002}$\\
Antiprotons & $0.010 \pm 0.003~^{+0.010}_{-0.004}$\\
\hline
Total & $0.107 \pm 0.032$ (stat $\oplus$ syst) \\

\end{tabular}
\end{table}

\section{Summary}
The Mu2e experiment will significantly improve the existing bounds in searching for a CLFV signal in  $\mu^-N - e^-N$ conversion. Mu2e is currently poised to begin Run 1 by the middle of the decade, with an expected 1000 times improvement on the existing world limit. Run 2, providing another factor of 10 improvement, is expected to begin by the end of the decade.

\section{Acknowledgements}
We are grateful for the vital contributions of the Fermilab staff and the technical staff of the participating institutions. This work was supported by the US Department of Energy (grant number 140000369); the Istituto Nazionale di Fisica Nucleare, Italy; the Science and Technology Facilities Council, UK; the Ministry of Education and Science, Russian Federation; the National Science Foundation, USA; the Thousand Talents Plan, China; the Helmholtz Association, Germany; and the EU Horizon 2020 Research and Innovation Program under the Marie Sklodowska-Curie Grant Agreement Nos. 101006726, 734303, 822185, and 858199. This document was prepared by members of the Mu2e Collaboration using the resources of the Fermi National Accelerator Laboratory (Fermilab), a U.S. Department of Energy, Office of Science, HEP User Facility. Fermilab is managed by Fermi Research Alliance, LLC (FRA), acting under Contract No. DE-AC02-07CH11359.

\bibliographystyle{elsarticle-num2}
\bibliography{Bibliography.bib}
\end{document}